\documentclass{IEEEtran}

\usepackage{style}
\usepackage[numbers, sort&compress]{natbib}
\usepackage{xr}
\usepackage{multicol}

\providecommand{\keywords}[1]
{
  \small	
  \textbf{\textit{Keywords---}} #1
}

\begin{document}

\title{How Can a Nation  Leverage  Quantum Technologies?}

\title{Why Should and How Can  a Nation  Leverage  Quantum Technologies?}

\title{Why Should Quantum Technologies Be Leveraged at National Levels and How?}

\title{Why Should and How Can Quantum Technologies Be Leveraged at National Levels?}
 
\author{AbdulMalek Baitulmal*, and  Nadia Adem**\\
     {*Quanta-ly
    Co., Tripoli, Libya, Email:a.baitulmal@quanta.ly}\\
    {**Department of Electrical and Electronic Engineering,
    University of Tripoli,  Libya,
    Email: n.adem@uot.edu.ly}\\

\large{Invited Paper}
}

\maketitle


\begin{abstract}
    Quantum technologies (QT) promise to change the landscape of technologies disruptively in diverse industries. For this reason, many nations around the globe are investing  to emerge {within}  the global quantum ecosystem through initiating national programs and international partnerships. Nonetheless, some other countries are still running behind and yet their governments need to take series actions to help their private and public sectors  adapt to the looming changes, considering the new regulations required and the huge influence that QT will present in the near future. 
    {In this opinion piece, we  provide, for the best of our knowledge, the first generally applicable, yet comprehensive and brief,  framework for leveraging the emerging quantum technologies to facilitate the establishment of national initiatives properly. 
    The insights presented in this article were driven based on investigating various approaches, initiatives, and roadmaps adopted globally and meeting  {with local and regional leaders, professionals, and governmental officials}. Furthermore, taken into account socioeconomic and institutional dimensions of the Libyan society, 
    we project the framework for the Libyan nation.  This opinion piece is intended to inspire researchers, technical industrial experts, stakeholders, and governmental bodies to find roles they need to play to bring QT forward.}\\  
 
\end{abstract}


\section{Introduction}
   The development of technologies has tremendously advanced as scientists gain a deeper insight about the way nature works. The theory of quantum mechanics has allowed us to design technologies that can improve various applications  including computation, communications, sensing and metrology, health, to name a few. 
   Thanks to many quantum properties; superposition, entanglement, teleportation, non-cloning, waves-interference and others, which have made quantum technologies (QT)  possible today~\cite{lopez2019quantum}.
   As the world is heading to the second quantum revolution,  preparing extensive strategic roadmaps and distinctive technology assessments to leverage QT are becoming inevitable. In the last decades, while there are many countries have set considerable efforts in this regard, see e.g.~\cite{rydlichowski19quantum, acin2018quantum,riedel2019europe,crane2017assessment}, many others are not yet prepared for  the revolution.
 
   To help accelerate the uptake of these countries, nevertheless, we, in this opinion piece, provide a  framework for leveraging QT.
   While taking into account the efforts that the regional and global leading countries are putting to establish proper quantum ecosystems, for example{~\cite{forbes2021toward,knight2019uk,raymer2019us}}, along with analyzing the Libyan landscape with a focus on the communications sector, we address  the  unique economical and social situations that help or challenge implementing the framework for the Libyan nation. 
   
   This article  raises the awareness of the need of QT implementation, emphasizes on the importance of the research and development in the transition to innovation and technology, analyzes the challenges of the investment landscapes, and discusses the necessity of the legislation in facilitating the implementation. Hence it can be found handy for a diverse body of readers ranging from learners who want to pick up the topic,   researchers driving the quantum uptake,  entrepreneurs bringing the technology forward, and governmental bodies  from leading countries  identifying partnership possibilities and other countries setting up the ground for  accelerating joining the quantum race.

    The rest of the paper is organized as follows: {Section~\ref{Potentials} highlights the potential of QT by shedding the light on the promising technical scenarios of the technology with a focus on the telecommunications sector.} Section~{\ref{Efforts}}, on the other hand, summarizes the  quantum efforts taken worldwide to build up an intuition of the global QT ecosystem.  {The necessity of acting  towards QT  is being discussed in Section~{\ref{Joining Crowd}}}.  Section~{\ref{Framework}} presents a potential framework for QT implementation. The section also discusses potentials and challenges of adapting such a framework in Libya. The comment piece is concluded in  Section~\ref{Conclusion}.

\section{QT Potentials}
\label{Potentials}
    Nowadays supercomputers, in spite of their strengths and advancements, are still unable to solve many complex  problems  in a reasonable amount of time. Mean while, a quantum computer processes  data in a parallel fashion, by leveraging quantum mechanics phenomena, giving rise to whole new aspect of computing algorithms.  
    {Currently}, researchers are working with industry leaders to develop quantum computing use cases to solve real-life  problems and meet business needs~\cite{hassija2020present}. To show the impact of QT, we   discuss quantum computing and  communications potentials in the telecommunications industry, in some details, and   for a number of other industry verticals, briefly. 
    
\subsection{Information and Communication Technologies {(ICT)}}
    The development of telecommunications networks from one generation to the next depends mainly on supporting end-user and network terminals demands. For the fifth generation of communications (5G), e.g., most of developed networking techniques came to prop complex services and  hugely improve  capacity and connection response between devices. Furthermore,  the sixth generation  (6G) will  be required to continue improving services, orders of magnitude, while being more efficient, for many potential new  use cases and exponentially increased number of users existing not only on ground, but also air, space, and underwater.
    Consequently,  there will be, incontrovertibly, need for automation, enabled mainly by artificial intelligence (AI), not only to meet the ever increasing demands of end-users but also to manage 6G networks,  expected to, for the first time, integrate various services such as sensing,  computing, positing along with communications~\cite{adem2021crucial}.
    Future networks  will require, to handle its complexity, new aspect of computing paradigms  which can  greatly benefit  from the power of quantum computing (QC) not only to speed up computations but also to handle traditionally unsolvable  problems~{\cite{adem2021crucial,botsinis2018quantum}. 
   
   Despite the potentials  QC offers to future networks, it, though, raises threats of breaking the existing end-to-end encryption schemes 
   which secure  data transmission through the ICT infrastructures. 
       Accordingly,  united states national institute of standards and technology (NIST) is involved in the development of post quantum cryptography (PQC) as new data-exchange encryption standards. These  schemes, technically handled as encryption certificates, are intended to provide the safe data transmission. Billions of old and new devices will need to transit to the new suite of certificates~\cite{joseph2022transitioning} as  NIST estimated to  approves them  in 2024~\cite{alagic2022status}.   

In addition to the efforts taken toward protecting infrastructures against QC threats, quantum mechanics  are   involved in securing optical communications, through for example the use of quantum key distribution (QKD) method.  QKD is considered as one of the building blocks of the future quantum internet by establishing quantum nodes~\cite{gyongyosi2022advances}. Therefore, facilitating the implementation of both quantum communications and QC networks, where services are envisioned to be powered by cloud-based quantum computers integrated with high-performance computing (HPC) infrastructure being secured by both PQC and QKD~\cite{cao2022evolution,ahn2022toward}.
     New cloud models, standards, and regulations are to be formulated for emerging services and deployments such as quantum computing as a service (QCaaS), quantum simulation, blind quantum computing, and distributed quantum computing systems leading to revolutionizing  the cloud computing architecture development.

 \subsection{Other QT potentials}
    QT have a wide range of potentials in many areas and fields. 
    Leveraging QC can take, for example,  business intelligence and logistics industries to the next level.      Leaders in these industries can perform complex decision making   in market prediction and mitigating investment risks~\cite{yndurain2019exploring}. QC power can also improve scheduling, planning for sophisticated business tasks, and  enhance traffic and congestion optimization~\cite{goddard2017will,gerbert2018next}. 
     {In addition, although QC is still restricted and limited to the pace of technical improvements  of  hardware, in  industries like  the chemical and drug developments,  researchers   discovered how to model quantum-mechanical systems on a much larger and complex scale than ever before. In the future, by using the duo quantum-AI, scientists in this field will have the ability to minimize severe time complexity when simulating large chemical and protein compounds~\cite{robert2021resource}}.

The research and development efforts~\cite{srivastava2016commercial,roberson2021talking} that are  being carried out to investigate QT use cases and applications in aforementioned and many other industries impacting many social and business aspects are leading to the development of global quantum ecosystem.  
A general overview of the QT efforts worldwide is presented in the next section.

\section{Existing Efforts}
\label{Efforts}

    Materialization of QT from academia and labs to market, in the last decade, required ever increasing funds. 
     {The total global quantum effort for  public funds and private investments for 2022 has been projected to be \$35.5b exceeding that couple of years ago by \$10b~\cite{global2022state}}.
    If one focuses on  nations landscape, the estimations put China as the leading country with  \$15b of investments, followed by Germany, France, US, UK, Canada, India each of which assign more than \$1b for their quantum initiatives~\cite{qureca2022overview}. Countries like Russia, Japan, Taiwan, Sweden, Austria, Singapore, on the other hand, each of which are investing  hundreds of millions~\cite{qureca2022overview}.
    Governments are funding to support preparing hardware,  setting up business incubation platforms for startups, establishing educational programs, growing talents, forming partnerships between academia and industries nationally and internationally. 
    The big increase in funding of national initiatives indicates the fast pace of forging the global ecosystem. 
    Accordingly this has stimulated private efforts  angel investors and venture capitalists (VCs) to join the quantum race resulting in  boosting  startups in many nations~\cite{gibney2019quantum,gerbert2018next,seskir2022landscape}.

    Despite all of these significant efforts, however,  many countries are still lagging behind with no  related noticeable investments or any clear roadmaps for leveraging QT. Such nations need to join the race now rather than later for reasons we discuss  in the next section.
  
\section{Crowd Joining}
\label{Joining Crowd}
 The enormous potentials and promises of QT are encouraging many countries to establish their own QT programmes to join the race influencing the development of the global ecosystem. However,  for many lagging nations, the opportunities to leverage such technologies are still undefined, which raises future entry barriers to the field. It is crucial for governments and organizations, nonetheless, in these countries to become early quantum adopters, for number of reasons. 

One of the critical conditions for QT is that they are fundamentally different from classical technologies, as each of these technologies has its own set of 
 implementation criteria. Thus, organizations have to develop their business models with new aspects and verities tailored to acquire QT solutions.  Similar to previous technology revolutions, QT  require developing the   mindsets that can envision their potentials and create roadmaps to facilitate their implementation. Nonetheless, efforts needed to get to that level of maturity can be enormous and very demanding in time and hence any related investments need to commence without further delay. 
Moreover,  forging  quantum roadmaps may face number of  hurdles, as we discuss later,  that are yet to be handled which could also be time consuming. 

Many QT potentials are yet to be discovered, nations acting in a timely manner do not only get the opportunity to join the crowd, but to also to influence the technology development and becoming a pioneer in identifying the industry needs contributing to {shaping future industries}.  In reality, as  use cases of QT may vary from a nation to another, countries do not involve in defining theirs are at the risk of being missed out. 

To facilitate having a nation joining the crowd and establish its corresponding intuitive, nonetheless, we, in the next section,   lay down the pillars of implementing a quantum framework.    

\section{Implementing a QT Framework}
\label{Framework}

    Due to the rapid and unprecedented technological developments in  recent years, many hurdles have emerged to governments that did not prepare roadmaps for digital transformation (DX).  Such a situation will massively reoccur in the future when quantum technologies are  commercialized globally. Many sectors must be included in the conversation, and suitable assessments must be made. This can be considered as quantum transformation (QX) on a national level. 
   In  this section, we provide a framework  for establishing a national quantum initiative. The framework is designed based on analyzing the global ecosystems and  challenges facing existing QT programmes. Furthermore, we analyze  obstacles that can arise to countries or organizations that have the intent to establish a QT roadmap. 
    The framework and its implementation are projected on the Libyan case taken into account global quantum strategists and  discussions made with local technology assessment leaders and governmental officials. The framework pillars discussed below are to be  all  or individually adopted depending on what suits the nation  implementing the technology.

\subsection{Workforce Preparation}  
    A big challenge which faces the adoption of the QT is the lack of skillful workforce specialized in diverse fields across the quantum landscape.
    
   Governments can lower this barrier through establishing dedicated quantum educational programmes and research projects, and build partnerships with universities and other focused entities regionally and globally. Some preliminary work is usually required beforehand though for example to identify the missing resources in the educational systems and have a government benefit  the most from partnerships. 

    Enterprises, on the other hand,  need special training programmes, where they can build a robust know-how about QT. This actually will help preparing the workforce that will be needed for various reasons such as maintaining and operating the technologies infrastructure.
    
    
\subsection{Environment Creation} 
    
    To facilitate the development of QT and their implementation, awareness among business leaders, professionals, and governmental officials,  among others need to be raised and quantum environment need to be built in some countries. Chances vary among these nations on how the environment is created, and which sectors are playing roles in doing so. For example,  with the increasing of the possibility of holding more events online,  the Libyan quantum community has started evolving quickly recently.  	In addition, that  facilitated Libyan experts, in fields including nanotechnology, quantum computers, data science, and ICT, from all over the globe engage in the environment  building. 
    Setting up a proper environment speeds up identifying the requirements for workforce preparation  {and hence} 
    defining the needs to adopt new technologies.
    
 \subsection{Needs Defining} 
  To establish a quantum initiative, nations must identify and understand the role of QT in local industries. In order to define the requirements for implementing QT, a number of milestones need to be considered.
   
  When organizations are adopting QT, they need to resonate with technologies, not  over expect from  or  under estimate them. This may be achieved by looking over the building blocks of classical technologies,  including AI, data analysis,  cybersecurity, etc,   to understand the limitations in  their capabilities and hence determine the real need for more advanced alternatives, mainly the QT.
        Thus, guaranteeing a starting point to avoid any QT hypes and deliver qualitative analysis about potential uses cases and technologies impact. 

 As a way to demonstrate quantum advantages over classical counterparts, initiatives must implement pilot projects. Such projects have a dominant role in creating use cases and understanding the role of technology in future infrastructures. As an example, in the aspect of quantum communications, ICT sectors can examine the PQC standards integration and verify leveraging the  existing optical fiber infrastructures with QKD solutions 
 and define their needs accordingly. 

    The importance of defining the needs for QT is that their corresponding solutions  are nation-oriented. Accordingly, governments must recognize the dimensions of their country in terms of the viable resources in industries, social impact, and their countries major 
    investments. Having their needs specified, nations can commence the transformation from the digital to the quantum paradigm.
           
 \subsection{Technology Transition}
    Unlike shallow technologies (e.g. digital technologies) which can be adopted and applied in many sectors without severe technical implementation difficulties, QT require different strategies to be implemented, as we discussed earlier. QT are similar, yet distinct, to other deep technologies (deep-tech) including AI, blockchain,  robotics,  etc.
    Such technologies depend fundamentally on advanced science, complex engineering, and creativity in product development.
  Hence, we believe that QX has many aspects of adoption to consider. 

   There are two types of quantum computing paradigms  universal quantum computing , and quantum annealing, which are referred to as digital and analog approaches respectively.    {Furthermore, there are various quantum computing physical  modalities  for {constructing} quantum processors, for example superconducting circuits, trapped-ions, quantum-dots, topological, and photonics based quantum processors.} Considering the varieties of existing implementation methods,  when building quantum ecosystems and preparing feasibility plans  the most appropriate techniques to approach and adopt need to be identified taken into account demands as well as available resources.
 Reliability and capability of the local educational institutions and big firms, for example, matter in preparing the workforce needed for the technology forging process. To make the transformation smooth  integrity between academia, industries, and businesses will be of a vital importance.
    
    {When it comes to profitability out of technology transformation investments, deep-tech generally have the potential to be very disruptive, however, they tend to take a long time to mature and become market-ready,  not to mention the grand upfront capital expenses (CapEx) may be required for developing and expanding related solutions}.  In countries, like Libya, where investors expect a quick return on investment (ROI) and depend on the direct profit, the  transformation process is expected to be slow as a result of  probably having them not making well investments in getting ready for going the quantum race.
There are a number of other aspects that could continue to hold QT investments  in lagging countries   including market risks  {as institutions have little to no investments key performance indicators (KPIs) knowledge and   global ecosystem activities awareness  both required to  discover their local market potentials and make related investments.}
The good news, though, number of these  barriers may be lowered down by technical demonstrations and proven scientific advancements, which, as it was the case for the global ecosystem~\cite{seskir2022landscape}, could encourage VCs to boost startups.  Generally speaking,  well investments in preparing the workforce, raising the awareness, and {defining needs}, can lead to a more smooth technology transition. In addition to that, laws and regulations are very crucial to construct the framework of an ecosystem and to enable the technology transition as we discuss next.

\subsection{Legislation Facilitating}
Technical projects  implemented  in  the public sector require not only a specific workflow between engaged parties, but also a well regulated legislation architecture  identifies the roles of each enterprise.
  Many countries had challenges, for example, with applying DX in public sectors, due to unformed regulations for technology deployment~\cite{wynn2021government}.  Lack of established laws and policies results into a chaos ecosystem, as it was the case with many projects implemented by many firms~\cite{orji2019international}.   Dilemmas that can come into sight include conflict of interest, hindering initiatives, miss understanding of the role of technology, and obstructing the framework of corporations.
    {The lack of regulations between  private and public sectors, in addition, may result in private enterprises failing to provide comprehensive solutions.} 
  Innovative private initiatives that contribute to the local ecosystem can end up shrinking, as a consequence. 
  
To avoid the aforementioned issues, a significant policy making efforts are being set in leading countries such as US, UK, China, Japan, Germany, France, and many others~\cite{kung2021quantum,riedel2019europe}. Countries leading in leveraging QT are facilitating regulations development such as QT use cases  can meet commercialization demands in their local markets through  using  existing infrastructures and establishing technology pilot projects. 

 However, governmental organizations and institutions, e.g. General Authority for Telecommunications, the Informatics  and National Information Security and Safety Authority (NISSA), the Libyan Post Telecommunications and Information Technology Holding Company (LPTIC), the Libyan Center for Strategic Studies and National Security in case of Libya, still need to act   setting up  policies, rules, and regulations and ensure their establishment. Thus, clearing up the way for  a QT initiative and preserving the feasibility of a national quantum roadmap which can be created throughout the cooperation between relevant legislative and executive authorities. 
%
\section{Conclusion}
\label{Conclusion}
	In this opinion piece, we presented QT  potentials in number of industries including the ICT. We also provided an overview of the global quantum efforts. 
    We discussed why
     nations need to make their strategic plans and define their  needs {for adopting  QT}. Moreover, we provided a framework that demonstrates basic pillars for  leveraging QT.   
    Most  QT potentials are yet to be uncovered and many uses cases need business mindsets to be created. This requires efforts from  enthusiasts to bring new aspects of leveraging QT.
    The opportunity is wide open for lagging countries to innovatively  join the crowd empowering their economy and influencing the global ecosystem.

\section*{Acknowledgement}
The authors would like to thank Abdulrahman Abudhair for   his   invaluable feedback about the projected framework, and   Saifeddin Hashad for the  great insights he has provided about developing  quantum initiatives in the Libyan ICT sector. The authors would like also  gratitude to André M. König, Farai Mazhandu, Brian Lenahan, and Rupesh Srivastava for their fruitful discussions about the quantum global ecosystem.

\bibliographystyle{unsrtnat}

\bibliography{refs}

\begin{thebibliography}{29}
\providecommand{\natexlab}[1]{#1}
\providecommand{\url}[1]{\texttt{#1}}
\expandafter\ifx\csname urlstyle\endcsname\relax
  \providecommand{\doi}[1]{doi: #1}\else
  \providecommand{\doi}{doi: \begingroup \urlstyle{rm}\Url}\fi

\bibitem[Lopez(2019)]{lopez2019quantum}
M~Allende Lopez.
\newblock Quantum technologies digital transformation, social impact, and
  cross-sector disruption.
\newblock \emph{Interamerican Development Bank}, 2019.

\bibitem[Rydlichowski et~al.()Rydlichowski, Naegele-Jackson, DFN, Jeannin,
  Chown, Golub, Vicinanza, Roberts, Vohnout, Skoda,
  et~al.]{rydlichowski19quantum}
Piotr Rydlichowski, Susanne Naegele-Jackson, Peter~Kaufmann DFN, Xavier
  Jeannin, Tim Chown, Ivana Golub, Domenico Vicinanza, Guy Roberts, Rudolf
  Vohnout, Pavel Skoda, et~al.
\newblock Quantum technologies status overview.
\newblock \emph{Quantum}, 19:\penalty0 01--2021.

\bibitem[Ac{\'\i}n et~al.(2018)Ac{\'\i}n, Bloch, Buhrman, Calarco, Eichler,
  Eisert, Esteve, Gisin, Glaser, Jelezko, et~al.]{acin2018quantum}
Antonio Ac{\'\i}n, Immanuel Bloch, Harry Buhrman, Tommaso Calarco, Christopher
  Eichler, Jens Eisert, Daniel Esteve, Nicolas Gisin, Steffen~J Glaser, Fedor
  Jelezko, et~al.
\newblock The quantum technologies roadmap: a european community view.
\newblock \emph{New Journal of Physics}, 20\penalty0 (8):\penalty0 080201,
  2018.

\bibitem[Riedel et~al.(2019)Riedel, Kovacs, Zoller, Mlynek, and
  Calarco]{riedel2019europe}
Max Riedel, Matyas Kovacs, Peter Zoller, J{\"u}rgen Mlynek, and Tommaso
  Calarco.
\newblock Europe’s quantum flagship initiative.
\newblock \emph{Quantum Science and Technology}, 4\penalty0 (2):\penalty0
  020501, 2019.

\bibitem[Crane et~al.(2017)Crane, Joneckis, Acheson-Field, Boyd, Corbin, Han,
  and Rozansky]{crane2017assessment}
Keith~W Crane, Lance~G Joneckis, Hannah Acheson-Field, Iain~D Boyd, Benjamin~A
  Corbin, Xueying Han, and Robert~N Rozansky.
\newblock \emph{Assessment of the Future Economic Impact of Quantum Information
  Science}.
\newblock JSTOR, 2017.

\bibitem[Forbes et~al.(2021)Forbes, Petruccione, and Roux]{forbes2021toward}
Andrew Forbes, Francesco Petruccione, and Filippus~S Roux.
\newblock Toward a quantum future for {s}outh {a}frica.
\newblock \emph{AVS Quantum Science}, 3\penalty0 (4):\penalty0 040501, 2021.

\bibitem[Knight and Walmsley(2019)]{knight2019uk}
Peter Knight and Ian Walmsley.
\newblock Uk national quantum technology programme.
\newblock \emph{Quantum Science and Technology}, 4\penalty0 (4):\penalty0
  040502, 2019.

\bibitem[Raymer and Monroe(2019)]{raymer2019us}
Michael~G Raymer and Christopher Monroe.
\newblock The us national quantum initiative.
\newblock \emph{Quantum Science and Technology}, 4\penalty0 (2):\penalty0
  020504, 2019.

\bibitem[Hassija et~al.(2020)Hassija, Chamola, Saxena, Chanana, Parashari,
  Mumtaz, and Guizani]{hassija2020present}
Vikas Hassija, Vinay Chamola, Vikas Saxena, Vaibhav Chanana, Prakhar Parashari,
  Shahid Mumtaz, and Mohsen Guizani.
\newblock Present landscape of quantum computing.
\newblock \emph{IET Quantum Communication}, 1\penalty0 (2):\penalty0 42--48,
  2020.

\bibitem[Adem et~al.(2021)Adem, Benfaid, Harib, and Alarabi]{adem2021crucial}
Nadia Adem, Ahmed Benfaid, Ramy Harib, and Anas Alarabi.
\newblock How crucial is it for 6{G} networks to be autonomous?
\newblock \emph{arXiv preprint arXiv:2106.06949}, 2021.

\bibitem[Botsinis et~al.(2018)Botsinis, Alanis, Babar, Nguyen, Chandra, Ng, and
  Hanzo]{botsinis2018quantum}
Panagiotis Botsinis, Dimitrios Alanis, Zunaira Babar, Hung~Viet Nguyen, Daryus
  Chandra, Soon~Xin Ng, and Lajos Hanzo.
\newblock Quantum search algorithms for wireless communications.
\newblock \emph{IEEE Communications Surveys \& Tutorials}, 21\penalty0
  (2):\penalty0 1209--1242, 2018.

\bibitem[Joseph et~al.(2022)Joseph, Misoczki, Manzano, Tricot, Pinuaga,
  Lacombe, Leichenauer, Hidary, Venables, and Hansen]{joseph2022transitioning}
David Joseph, Rafael Misoczki, Marc Manzano, Joe Tricot, Fernando~Dominguez
  Pinuaga, Olivier Lacombe, Stefan Leichenauer, Jack Hidary, Phil Venables, and
  Royal Hansen.
\newblock Transitioning organizations to post-quantum cryptography.
\newblock \emph{Nature}, 605\penalty0 (7909):\penalty0 237--243, 2022.

\bibitem[Alagic et~al.(2022)Alagic, Apon, Cooper, Dang, Dang, Kelsey,
  Lichtinger, Miller, Moody, Peralta, et~al.]{alagic2022status}
Gorjan Alagic, Daniel Apon, David Cooper, Quynh Dang, Thinh Dang, John Kelsey,
  Jacob Lichtinger, Carl Miller, Dustin Moody, Rene Peralta, et~al.
\newblock Status report on the third round of the nist post-quantum
  cryptography standardization process.
\newblock \emph{US Department of Commerce, NIST}, 2022.

\bibitem[Gyongyosi and Imre(2022)]{gyongyosi2022advances}
Laszlo Gyongyosi and Sandor Imre.
\newblock Advances in the quantum internet.
\newblock \emph{Communications of the ACM}, 65\penalty0 (8):\penalty0 52--63,
  2022.

\bibitem[Cao et~al.(2022)Cao, Zhao, Wang, Zhang, Ng, and
  Hanzo]{cao2022evolution}
Yuan Cao, Yongli Zhao, Qin Wang, Jie Zhang, Soon~Xin Ng, and Lajos Hanzo.
\newblock The evolution of quantum key distribution networks: On the road to
  the qinternet.
\newblock \emph{IEEE Communications Surveys \& Tutorials}, 24\penalty0
  (2):\penalty0 839--894, 2022.

\bibitem[Ahn et~al.(2022)Ahn, Kwon, Ahn, Park, Kim, Lee, Kim, and
  Chung]{ahn2022toward}
Jongmin Ahn, Hee-Yong Kwon, Bohyun Ahn, Kyuchan Park, Taesic Kim, Mun-Kyu Lee,
  Jinsan Kim, and Jaehak Chung.
\newblock Toward quantum secured distributed energy resources: Adoption of
  post-quantum cryptography ({PQC}) and quantum key distribution ({QKD}).
\newblock \emph{Energies}, 15\penalty0 (3):\penalty0 714, 2022.

\bibitem[Yndurain et~al.(2019)Yndurain, Woerner, and
  Egger]{yndurain2019exploring}
Elena Yndurain, Stefan Woerner, and Daniel~J Egger.
\newblock Exploring quantum computing use cases for financial services.
\newblock \emph{IBM Institute for business value. URL: https://www. ibm.
  com/downloads/cas/2YPRZPB3.[dl: 29.07. 2020]}, 2019.

\bibitem[Goddard et~al.(2017)Goddard, Mniszewski, Neukart, Pakin, and
  Reinhardt]{goddard2017will}
Phil Goddard, Susan Mniszewski, Florian Neukart, Scott Pakin, and Steve
  Reinhardt.
\newblock How will early quantum computing benefit computational methods?
\newblock In \emph{Proc. SIAM Annu. Meeting}, 2017.

\bibitem[Gerbert and Rue{\ss}(2018)]{gerbert2018next}
Philipp Gerbert and Frank Rue{\ss}.
\newblock The next decade in quantum computing and how to play.
\newblock \emph{Boston Consulting Group}, page~5, 2018.

\bibitem[Robert et~al.(2021)Robert, Barkoutsos, Woerner, and
  Tavernelli]{robert2021resource}
Anton Robert, Panagiotis~Kl Barkoutsos, Stefan Woerner, and Ivano Tavernelli.
\newblock Resource-efficient quantum algorithm for protein folding.
\newblock \emph{npj Quantum Information}, 7\penalty0 (1):\penalty0 1--5, 2021.

\bibitem[Srivastava et~al.(2016)Srivastava, Choi, Cook, Team,
  et~al.]{srivastava2016commercial}
Rupesh Srivastava, Iris Choi, Tim Cook, NUE Team, et~al.
\newblock The commercial prospects for quantum computing.
\newblock \emph{Networked Quantum Information Technologies}, 2016.

\bibitem[Roberson et~al.(2021)Roberson, Leach, and Raman]{roberson2021talking}
Tara Roberson, Joan Leach, and Sujatha Raman.
\newblock Talking about public good for the second quantum revolution:
  analysing quantum technology narratives in the context of national
  strategies.
\newblock \emph{Quantum Science and Technology}, 6\penalty0 (2):\penalty0
  025001, 2021.

\bibitem[on~Quantum~Computing(2022)]{global2022state}
Global Future~Council on~Quantum~Computing.
\newblock State of quantum computing: Building a quantum economy.
\newblock World Economic Forum, 2022.

\bibitem[Venegas-Gomez(2022)]{qureca2022overview}
Araceli Venegas-Gomez.
\newblock Overview on quantum initiatives worldwide.
\newblock \emph{QURECA Blog.
  URL:https://qureca.com/overview-on-quantum-initiatives-worldwide-update-2022/.},
  2022.

\bibitem[Gibney(2019)]{gibney2019quantum}
Elizabeth Gibney.
\newblock The quantum gold rush.
\newblock \emph{Nature}, 574\penalty0 (7776):\penalty0 22--24, 2019.

\bibitem[Seskir et~al.(2022)Seskir, Korkmaz, and
  Aydinoglu]{seskir2022landscape}
Zeki~Can Seskir, Ramis Korkmaz, and Arsev~Umur Aydinoglu.
\newblock The landscape of the quantum start-up ecosystem.
\newblock \emph{arXiv preprint arXiv:2205.01999}, 2022.

\bibitem[Wynn et~al.(2021)Wynn, Bakeer, and Forti]{wynn2021government}
Martin~G Wynn, Ali Bakeer, and Yousef Forti.
\newblock E-government and digital transformation in {L}ibyan local
  authorities.
\newblock \emph{International Journal of Teaching and Case Studies},
  12\penalty0 (2):\penalty0 119--139, 2021.

\bibitem[Orji(2019)]{orji2019international}
Uchenna~Jerome Orji.
\newblock \emph{International Telecommunications Law and Policy}.
\newblock Cambridge Scholars Publishing, 2019.

\bibitem[Kung and Fancy(2021)]{kung2021quantum}
Johnny Kung and Muriam Fancy.
\newblock A quantum revolution: Report on global policies for quantum
  technology.
\newblock \emph{CIFAR, Toronto}, 2021.

\end{thebibliography}

\end{document}